\documentclass[twocolumn]{aastex631}
\usepackage{booktabs}
\usepackage{eqnarray,amsmath}

\hypersetup{linkcolor=blue,citecolor=blue,filecolor=cyan,urlcolor=blue}

\begin{document}

\title{Rings Around Non-Spherical Worlds\\ Sub-mm Dust Retention Around Triaxial Small Bodies in the Solar System}

\correspondingauthor{Zsolt Regály}
\email{regaly@konkoly.hu}
\author[0000-0001-5573-8190]{Zsolt Regály}
\affiliation{HUN-REN CSFK Konkoly Observatory, MTA Centre of Excellence, Konkoly Thege M. \'ut 15-17, Budapest, 1121, Hungary}

\author[0000-0003-3780-7185]{Viktória Fröhlich}
\affiliation{HUN-REN CSFK Konkoly Observatory, MTA Centre of Excellence, Konkoly Thege M. \'ut 15-17, Budapest, 1121, Hungary}
\affiliation{ELTE Eötvös Loránd University, Institute of Physics and Astronomy, P\'azm\'any P\'eter S\'et\'any 1/A, Budapest, 1117, Hungary}

\author[0000-0002-1663-0707]{Csilla Kalup}
\affiliation{HUN-REN CSFK Konkoly Observatory, MTA Centre of Excellence, Konkoly Thege M. \'ut 15-17, Budapest, 1121, Hungary}
\affiliation{ELTE Eötvös Loránd University, Institute of Physics and Astronomy, P\'azm\'any P\'eter S\'et\'any 1/A, Budapest, 1117, Hungary}

\author[0000-0002-8722-6875]{Csaba Kiss}
\affiliation{HUN-REN CSFK Konkoly Observatory, MTA Centre of Excellence, Konkoly Thege M. \'ut 15-17, Budapest, 1121, Hungary}
\affiliation{ELTE Eötvös Loránd University, Institute of Physics and Astronomy, P\'azm\'any P\'eter S\'et\'any 1/A, Budapest, 1117, Hungary}

\begin{abstract}
We investigated the millennial-scale evolution of narrow innermost rings composed of pebble-sized to sub-millimeter particles around the four known ring-bearing small bodies Chiron, Chariklo, Quaoar, and Haumea. 
Using a GPU-accelerated fourth-order Hermite integrator, we modeled the combined effects of solar radiation pressure, shadowing of the rings by the host body, heliocentric motion, and the non-axisymmetric gravitational field of the rotating triaxial central body. 
The calculations compare spherical and triaxial-body models, as well as coplanar and inclined ring configurations. 
In spherical models, solar radiation pressure efficiently excites particle eccentricities, leading to accretion onto the central body above a critical radiation-pressure parameter. 
This effect is strongest for the lower-mass systems, Chiron and Chariklo, where particles with relatively modest radiation forcing are rapidly removed. 
In contrast, when the triaxial shape of the host body is included, rapid apsidal precession suppresses radiation-pressure-driven eccentricity growth and prevents material loss from the ring over the simulated interval. 
The triaxial models also suppress the previously identified Sun-facing reorientation of highly inclined rings and instead produce moderate vertical broadening.
Strongly confined rings persist for radiation-pressure parameters corresponding to particle sizes larger than about $7–40~\mu$m, depending on composition. 
Their characteristic radial widths are about $10$~km for Chiron and Chariklo and about $40-70$~km for Quaoar and Haumea.
The vertical thicknesses of the rings are estimated to be on the order of 1~km for Chiron and Chariklo, and only several hundred meters for Quaoar and Haumea.
Our results therefore suggest that narrow rings around triaxial small bodies in the Solar System can plausibly retain sub-millimeter particles over dynamically relevant timescales shorter than Poynting-Robertson drag.
\end{abstract}

\keywords{small body rings (1254) --- Dwarf planets (419) --- Trans-Neptunian objects (1705)}

\section{Introduction}

During the last decade, occultation measurements have revealed that rings are not only hosted by giant planets. 
In 2013, the first discovery of a double ring system around the Centaur-type minor planet (10199)~Chariklo \citep{BragaRibas2014Chariklo} generated a large number of dynamical studies and drew more attention to occultation campaigns. 
In 2017 and 2023, rings were also found in the trans-Neptunian region around the dwarf planets (136108)~Haumea \citep{Ortiz2017Haumea} and (50000)~Quaoar \citep{Morgado2023,Pereira2023QuaoarRings}. 
Recently, another Centaur, the long-suspected (2060)~Chiron, also revealed evidence for a rapidly evolving ring system showing complex structures \citep{Ortiz2015Chiron,Ortiz2023,Pereira2025ChironRings}.

While the different dynamical environments and the unique properties of the main bodies can vary widely, there are several remarkable similarities, such as the presence of multiple rings, and the fact that rings are usually close to the 1:3 spin–orbit resonance (SOR) with the main body. 
To date, the origin of these rings remains an open question, although several theories have been proposed for each system \citep{Sicardy2025Origin}. 
In order to evaluate these theories, obtaining information on the composition of the rings is crucial, as particle size distribution and possible constituent materials not only shed light on the origin of the rings but also strongly affect their long-term fate.

Due to their proximity to the central body, spectroscopic observations cannot resolve the rings themselves, allowing only integrated measurements. 
However, by taking into account the changing aspect angle of the rings, the dominant materials can be inferred, such as water ice in the case of Chariklo \citep{BragaRibas2014Chariklo,Duffard2014Chariklo}. 
Currently, the only observational constraint on the particle sizes comes from simultaneous occultation observations in two distinct passbands. 
For Chariklo, \cite{Morgado2021Chariklo} found no significant difference between the visual ($0.45-0.65~\mu$m) and red ($0.70-1.0~\mu$m) channels, indicating that the particle size must be larger than a few microns. 
However, dynamical simulations have so far considered only cm- to m-sized particles, i.e. pebbles. 
\cite{PanWu2016Chariklo} assumed that the suggested finite ring eccentricity and apsidal alignment of Chariklo's rings are maintained by the self-gravity of the ring, which yields a typical particle size of a few meters. 
However, \cite{Michikoshi2017Chariklo} found that N-body simulations of the self-gravitating collisional particle rings imply smaller ring particles or the existence of shepherding satellites. 
Shepherd moons have also been invoked by several other studies to explain the confinement and sharp edges of the observed rings \citep{GiuliattiWinter2023,Sickafoose2024}, as well as the origin of the rings themselves \citep{Melita2017Chariklo}.

Although several studies have favored direct analogies with giant planet ring systems (such as meter-sized pebbles in the ring), the observational evidence is not that strong. 
In contrast to giant planets, the gravitational field around irregular bodies introduces large non-axisymmetric terms that have a drastic influence on ring dynamics \citep{Sicardy2025}. 
This dynamical role of non-axisymmetric gravity around small bodies is now well established. 
Even in systems not primarily discussed in the context of rings, the topology of the phase space already depends sensitively on body shape and spin.
The locations and sizes of stability regions around irregular bodies vary strongly from case to case, with distinct behaviors for the dog-bone asteroid (216)~Kleopatra, the slowly rotating comet 9P/Tempel~1, and the primary of the triple near-Earth system 2001~SN263. 
For Haumea and Chariklo, the equatorial ellipticity shifts SORs, modifies their widths, and reshapes the extent of stable and chaotic regions around the central body \citep{Winter2019Haumea,Ribeiro2023}.

For the dwarf planet Haumea, \citet{Ribeiro2023} showed that the nominal 1:3 SOR is actually a doubled 2:6 SOR, with highly eccentric periodic orbits bounded by a chaotic separatrix. The ring is therefore more likely associated with first-kind periodic orbits than with strict resonant confinement. A similar result was found for Chariklo, where 1:3 resonant trajectories exceed the ring width, while first-kind periodic orbits remain consistent with the observed ring region \citep{Madeira2022}. In the triaxial Chariklo model of \citet{GiuliattiWinter2023}, the stable region is likewise shaped by first-kind and resonant families: C1R overlaps the first-kind stable corridor for suitable eccentricities, whereas C2R lies in a less favorable zone and may need additional confinement. The same trend holds for Chiron, where \citet{Madeira2025Chiron} showed that higher elongation expands the chaotic inner region and shifts the stable boundary outward, while lower elongation allows the disk to remain stable closer to the primary. Under nominal parameters, Chi2R is the best candidate for association with the bifurcated 1:3 SOR, whereas Chi1R is unlikely to be linked to the unstable 1:2 SOR.

Centaurs often show cometary-like activity that releases dust particles down to the submicron regime. In the case of Chiron, the last outburst occurred in 2021 \citep{Ortiz2023}, although the interaction of the ejected material with the ring is unclear \citep{Pereira2025ChironRings}, but might be a possible source. 
Small dust particles in planetary rings can also be continuously replenished through a collisional cascade, in which larger parent bodies or moonlets fragment through mutual collisions, and progressively grind down to micron-sized particles \citep{Esposito1993}. Such cascades are thought to operate in several Solar System ring systems, where the disruption of small satellites or embedded moonlets produces debris that subsequently evolves collisionally into fine dust. In addition, impacts onto nearby shepherd moons or ring-embedded bodies can eject material that feeds dusty rings \citep{Colwell2000}. The recent discovery of disk-like structures around Chiron \citep{Pereira2025ChironRings}, and the likely complex satellite--ring system around Quaoar \citep{Proudfoot2025,BragaRibas2026}, strongly indicate that such processes may also be active around smaller bodies. 
While most rings of the gas giants lack sub-millimeter particles \citep{Harbison2013Saturn,Nicholson2018Uranus}, notable exceptions include Jupiter’s main ring, halo, and gossamer rings, which are well known to be dust-rich and dominated by micron-sized grains.
Also Uranus' $\lambda$ and other dusty rings \citep{DePater2006Uranus,DePater2007Uranus} and Saturn's Phoebe ring \citep{Hamilton2015Phoebe,Tamayo2016} are dominated by small particles, down to micron- or even submicron-sized particles.

The first dedicated analysis of sub-millimeter particles in small-body rings was carried out by \cite{Kalup2024}, where they investigated different materials and particle sizes to characterize the possible spectral energy distributions of the ring system of Haumea. 
Comparing these models with future multiwavelength measurements offers a diagnostic tool for compositional analyses of rings. 
Recently, \cite{Kiss2024} reported an unexpected mid-infrared excess around another dwarf planet, Makemake, a feature predicted by \cite{Kalup2024}. 
One plausible explanation is a ring composed of very small carbonaceous grains, raising the need for dynamical studies of dust particles.
Previously, \cite{Regaly2025A&A...697A.116R} explored the evolution of rings around spherical Chariklo and Haumea analogs, taking into account the gravitational effects of the central body, the Sun, radiation pressure, and the shadow cast by the central body, and showed a radiation-pressure-driven configuration of highly inclined rings, the so-called sunflower effect.

In this paper, we continue these investigations on rings around Solar System small bodies by modeling the millennial-scale dynamical evolution of pebble-sized and sub-millimeter dusty rings around Chiron, Chariklo, Quaoar, and Haumea. 
Our primary objective is to assess the coupled effects of solar radiation pressure and the rotating triaxial gravitational potential of the central body on ring stability and morphology.

The paper is organized as follows. 
Section~2 presents a detailed description of the numerical model and underlying assumptions. 
In Section~3, we report the results of the simulations. 
Section~4 discusses the implications of these findings. 
Finally, Section~5 summarizes our conclusions.

\section{Numerical models}

We modeled ring dynamics using a high-precision, GPU-accelerated fourth-order Hermite integrator following the scheme of \citet{Makino1992PASJ...44..141M}. 
 Our numerical integrations are performed with the \textsc{HIPERION}\footnote{HIPERION (HIgh PERformance Integrator for N-body) is a GPU-accelerated, in-house developed, versatile N-body code. It is designed to model planet–disk interactions and to perform planet population synthesis calculations. The code employs 4th, 6th, and 8th-order Hermite integration schemes, and includes physical processes such as collisions between bodies, planetary accretion, radiation pressure, and Poynting–Robertson drag acting on massless dust particles. HIPERION has been successfully applied in numerous studies over the past decade (see, e.g., \citep{Regalyetal2018MNRAS.473.3547R,Regaly2023A&A...677L...6R,Regaly2025A&A...697A.116R,Dencs2019MNRAS.487.2191D,Dencs2021A&A...645A..65D,Dencs2025A&A...699A.166D}.} code.
The code employs a barycentric inertial coordinate system, which is required in the present context in order to consistently incorporate both solar radiation pressure and the gravitational potential of a rotating triaxial central body. 
Because radiation pressure is defined in the inertial frame and the non-axisymmetric ($C_{22}$) component of the gravity field rotates with the body, the commonly used equations of motion formulated in a purely body-fixed frame \citep{Hu2004P&SS...52..685H} are not directly applicable here.

In the following, we present I.) the algorithms used to compute the gravitational force of a rotating triaxial body, II.) our method for initializing an osculating circular ring, III.) the treatment of solar radiation pressure taking into account the shadow of the central body, and IV.) the physical properties of the investigated ring systems.

\subsection{Rotating Triaxial Gravity Field}
\label{sec:triaxial_hermite}

To represent a triaxial body, we model the central object as a point mass plus the dominant second order gravity harmonics, retaining the zonal $J_2$ and the real tesseral $C_{22}$, and assuming a rigid figure rotating at a constant spin rate $\Omega$ about the inertial $z$ axis.  
Using an equivalent radius $R_\mathrm{eq}$ and gravitational parameter $\mu_\mathrm{b}\equiv GM_\mathrm{b}$, where $G$ is the Newtonian gravitational constant and $M_\mathrm{b}$ the mass of the central body, we adopt the second-order potential at the body-centric position $\mathbf{r}_\mathrm{b}$ 
\begin{align}
V(\mathbf{r_\mathrm{b}})
  &= \frac{\mu_\mathrm{b}}{r_\mathrm{b}}
     \left(\frac{R_\mathrm{eq}}{r_\mathrm{b}}\right)^2
     \Bigg[1+
        \frac{1}{2}J_2
        \left(3(\hat{\mathbf{z}}\cdot\hat{\mathbf{r}}_\mathrm{b})^2-1\right)
\notag
\\
  &\qquad
        - \frac{3}{2}C_\mathrm{22}
        \left(
            (\hat{\mathbf{x}}\cdot\hat{\mathbf{r}}_\mathrm{b})^2
            -
            (\hat{\mathbf{y}}\cdot\hat{\mathbf{r}}_\mathrm{b})^2
        \right)
     \Bigg],
     \label{eq:pot}
\end{align}
where $R_\mathrm{eq}$ is the volumetric equivalent radius of the central body, $\hat{\mathbf{x}},~\hat{\mathbf{y}},~\hat{\mathbf{z}}$ are the Cartesian unit vectors, $\hat{\mathbf{r}}_\mathrm{b}$ is the unit vector towards $\mathbf{r}_\mathrm{b}$, and $r_\mathrm{b}=\|\mathbf{ r_b}\|$.
The triaxial shape is given as axis ratios (a, b, c)/a; thus, the normalized zonal and tesseral harmonic components are 
\begin{equation}
    J_{2}=\frac{a^2+b^2-2c^2}{10a},\quad C_{22}=\frac{a^2-b^2}{20a}.
\end{equation}

The fourth-order Hermite predictor--corrector scheme requires both the acceleration $\mathbf{a}=\dot{\mathbf{v}}$ and its first time-derivative (jerk) $\mathbf{j}=\dot{\mathbf{a}}$ at each force evaluation \citep{Makino1992PASJ...44..141M}.
The corresponding central body frame acceleration $\mathbf{a}_\mathrm{b}=\nabla V(\mathbf{r}_\mathrm{b})$ can be written as
\begin{equation}
\mathbf{a}_\mathrm{b} = -\mu_\mathrm{b}\frac{\mathbf{r_\mathrm{b}}}{r_\mathrm{b}^3}+\mathbf{a_\mathrm{J2}} + \mathbf{a}_\mathrm{C22}.
\label{eq:vect_a}
\end{equation}
In the body frame the gravity field is time-independent, hence the jerk is obtained from the Jacobian of the acceleration,
\begin{equation}
\mathbf{j}_\mathrm{b} = \frac{\partial \mathbf{a}_\mathrm{b}}{\partial \mathbf{r}_\mathrm{b}}~\mathbf{v}_\mathrm{b}=-\mu_\mathrm{b}\left(\frac{\mathbf{v}_\mathrm{b}}{r_\mathrm{b}^3}-3\frac{\mathbf{r}_\mathrm{b}\cdot\mathbf{v}_\mathrm{b}}{r_\mathrm{b}^5}\right) + \mathbf{j_{J2}}+ \mathbf{j_{C22}},
\label{eq:vect_j}
\end{equation}
where $\mathbf{v}_\mathrm{b}=\mathbf{\dot{r}}_\mathrm{b}$ is the particle's velocity relative to the central body.
The complete Cartesian expressions for potential, acceleration and jerk terms are given in Appendix~\ref{apx:acc_jerk}.

To model rigid rotation (with a constant angular speed $\Omega$), a transformation between inertial and body frames is required.
Let the body-to-inertial rotation be $Q(t)=\mathcal{R}_z(\theta)$ with $\theta(t)=\theta_0+\Omega t$, where $\mathcal{R}_z(\theta)$ is the rotational matrix. 
For an inertial particle state $(\mathbf{r}',\mathbf{v}')$ relative to the body's center,
\begin{equation}
\mathbf{r}_\mathrm{b} = Q^{\mathsf T}\mathbf{r}',
\quad
\mathbf{v}_\mathrm{b} = Q^{\mathsf T}\mathbf{v}' - \mathbf{\Omega}\times \mathbf{r}'_\mathrm{b},
\quad \mathbf{\Omega}=\Omega~\hat{\mathbf{z}}.
\label{eq:rbvb}
\end{equation}
We evaluate Eqs.~\eqref{eq:vect_a}--\eqref{eq:vect_j} in the central body frame and rotate back to inertial space via
\begin{equation}
\mathbf{a} = Q~\mathbf{a}_\mathrm{b}.
\label{eq:a_inertial}
\end{equation}

The inertial jerk must include the explicit time-dependence introduced by the rotating basis:
\begin{equation}
\mathbf{j} = Q~\mathbf{j}_\mathrm{b} + \mathbf{\Omega}\times \mathbf{a}.
\label{eq:j_inertial}
\end{equation}
Equation~\eqref{eq:j_inertial} is essential for maintaining fourth-order accuracy in the Hermite predictor--corrector when tesseral terms rotate in inertial space \citep{Makino1992PASJ...44..141M}.

To embed in an $N$-body Hermite force loop when the triaxial central body translates (e.g. heliocentric orbit), the aspherical contribution is evaluated using relative coordinates and velocities of the given ring particle at each force call. 
In practice, we compute the total $(\mathbf{a},\mathbf{j})$ as the sum of (i) pairwise Newtonian monopoles and (ii) a triaxial correction (Eqs.~\ref{eq:aJ2}--\ref{eq:j_inertial}) added to particles orbit the central body. This fits naturally into the Hermite predict--evaluate--correct pattern \citep{Makino1992PASJ...44..141M}.

\subsection{Force-consistent Osculating Circular Initialization}

For the initial conditions, we adopt an osculating circular-state prescription to construct a circular ring in the equatorial plane ($z_\mathrm{b}=0$) of the body-centric frame. 
For each particle at position $\mathbf{r}_\mathrm{b}$ at \(t_0=0\), we compute the inertial acceleration $\mathbf{a}(t_0,\mathbf{r}_\mathrm{b})$ using the same gravity routine as that employed by the numerical integrator. 
Its radial component,
\begin{equation}
a_r=\mathbf{a}\cdot\hat{\mathbf{r}}_\mathrm{b},
\end{equation}
is then used to define the osculating circular speed,
\begin{equation}
v_0=\sqrt{r_{\mathrm b}(-a_r)}.
\end{equation} 
The initial velocity vector is assigned in the tangential direction as
\begin{equation}
\mathbf{v}_0 = v_0 \left(\hat{\mathbf z}\times \hat{\mathbf r}_{\mathrm b}\right).
\end{equation}
The central body’s heliocentric orbital velocity must be added to $\mathbf{v}_0$.
This construction ensures consistency between the initialized state and the adopted force model, while avoiding ambiguities associated with normalization or sign conventions. 

For \(C_{22}\neq 0\), however, the resulting circularity is only instantaneous, since the non-axisymmetric potential introduces tangential forcing away from the principal axes. 
Consequently, strictly circular rings cannot remain long-lived in the rotating potential. 
Instead, a small eccentricity is excited in the ring particles, so that the ring becomes slightly eccentric (see Section~\ref{sec:discussion} for details).

\subsection{Solar Radiation Pressure}
\begin{deluxetable*}{lccccccccc}
\tablecaption{Physical and ring parameters for selected ringed small bodies and small bodies used in this study.\label{tab:ringed_bodies}}
\tablewidth{0pt}
\tabletypesize{\footnotesize}
\tablehead{
\colhead{Object} &
\colhead{$a$ (AU)} &
\colhead{$e$} &
\colhead{$R_{\rm eq}$ (km)} &
\colhead{$M$ (kg)} &
\colhead{$(a,b,c)/a$} &
\colhead{$r_{\rm ring}$ (km)} &
\colhead{$P_{\rm spin}$ (hr)} &
\colhead{$P_{\rm orb}$ (hr)} &
\colhead{$\Delta i$ (deg)}
}
\startdata
Chiron\tablenotemark{\small a}    & 13.70 & 0.38 & 98    &
$4.8\times10^{18}$   & $(1,~0.865,~0.540)$ & $325$  & 5.918 & 18.07 & 57.4 \\
Chariklo\tablenotemark{\small b}  & 15.74 & 0.17 & 143.8 &
$6.4\times10^{18}$   & $(1,
0.940,
~0.689)$ & $385.9$  & 7.004 & 20.65 & 57.9 \\
Quaoar\tablenotemark{\small c}    & 43.69 & 0.04 & 545   &
$1.4\times10^{21}$   & $(1,~0.840,~0.724)$ & $2520$ & 17.76 & 22.84 & 20.5 \\
Haumea\tablenotemark{\small d}    & 43.12 & 0.19 & 797.5 &
$4.006\times10^{21}$ & $(1,~0.734,~0.442)$ & $2287$ & 3.915 & 11.67 & 87.2 \\
\enddata
\tablecomments{$R_{\rm eq}$ is the volume-equivalent radius of the central body. The triaxial shape is given as axis ratios $(a,b,c)/a=(1,~b/a,~c/a)$ using the adopted ellipsoid model in the cited references. Ring distances $r_{\rm ring}$ are measured from the body center. Several of the small bodies possess multiple rings, with the exception of Haumea. In these cases, we modeled only the innermost ring. $P_\mathrm{spin}$ is the spin period of the small body. $P_{\rm orb}$ is the circular Keplerian orbital period at the adopted ring radius, computed from the listed mass as $P_{\rm orb}=2\pi(r_{\rm ring}^3/(GM))^{1/2}$.}
\tablenotetext{a}{\cite{Pereira2025ChironRings,BragaRibas2023Chiron,MarcialisBuratti1993Chiron}}
\tablenotetext{b}{\citet{Leiva2017Chariklo,BragaRibas2014Chariklo,Fornasier2014Chariklo}}
\tablenotetext{c}{\cite{Kiss2024QuaoarLightcurve,Pereira2023QuaoarRings,Fraser2013QuaoarWeywot}}
\tablenotetext{d}{\cite{Ortiz2017Haumea,RagozzineBrown2009Haumea}}
\label{tab:systems}
\end{deluxetable*}

\label{sec:radpress_hermite}

Radiation pressure provides (RP) a conservative, radially outward perturbation to the heliocentric motion of small particles, and it is an effective reduction of solar gravity. 
Using $\mu_\odot \equiv GM_\odot$, the radiation-pressure acceleration is
\begin{equation}
\mathbf{a}_{\rm RP}= \beta~\mu_\odot~\frac{\mathbf{r}}{r^3},
\label{eq:arp}
\end{equation}
where ${\boldsymbol r}$ is the heliocentric position vector of the particle and $\beta$ is the ratio of radiation pressure to solar gravity \citep{Burns1979Icar...40....1B}. 
Thus, the heliocentric equation of motion for a particle subject to gravity and radiation pressure can be written as
\begin{equation}
\ddot{\mathbf{r}} = -\mu_\odot\frac{\mathbf{r}}{r^3} + \mathbf{a}_{\rm RP}
= -(1-\beta)~\mu_\odot~\frac{\mathbf{r}}{r^3},
\label{eq:mu_eff}
\end{equation}
i.e. radiation pressure reduces the effective solar gravity to $(1-\beta)\mu_\odot$ \citep{Burns1979Icar...40....1B}. For nearly Keplerian motion this implies an effective mean motion
\begin{equation}
n^2 = \frac{(1-\beta)\mu_\odot}{a^3},
\label{eq:n_beta}
\end{equation}
and therefore shifts resonance locations and secular timescales relative to purely gravitational dynamics.

The fourth-order Hermite scheme requires both acceleration and jerk \citep{Makino1992PASJ...44..141M}. Differentiating Eq.~(\ref{eq:arp}) yields the radiation-pressure jerk
\begin{equation}
\mathbf{j}_{\rm RP}\equiv \dot{\mathbf{a}}_{\rm RP}
= \beta~\mu_\odot\left(\frac{\mathbf{v}}{r^3}
-3~\frac{(\mathbf{r}\cdot\mathbf{v})}{r^5}~\mathbf{r}\right),
\label{eq:jrp}
\end{equation}
which is algebraically identical to the Newtonian jerk but with opposite sign and scaled by $\beta$.

In the $N$-body Hermite force loop at each force evaluation, we compute heliocentric relative vectors for each particle $i$:
\begin{equation}
\mathbf{r} \equiv \mathbf{r}_\mathrm{i}-\mathbf{r}_\odot,\quad
\mathbf{v} \equiv \mathbf{v}_\mathrm{i}-\mathbf{v}_\odot.
\end{equation}
Radiation pressure can then be incorporated in either of two equivalent Hermite-consistent ways:

(1) Effective-$\mu$ (recommended when only RP is present): replace the solar contribution for each particle by $(1-\beta_i)\mu_\odot$ in the particle--Sun interaction Eq.~(\ref{eq:mu_eff}).
This approach is convenient when $\beta_i$ differs among particles.

(2) Explicit additive correction:
add $\mathbf{a}_{\rm RP}$ in Eq.~(\ref{eq:arp}) and $\mathbf{j}_{\rm RP}$ in Eq.~(\ref{eq:jrp}) directly to the particle's total $(\mathbf{a}_\mathrm{i},\mathbf{j}_\mathrm{i})$ after computing gravitational contributions. This is operationally similar to adding the triaxial correction and is straightforward when radiation forces are treated as external (non-action--reaction) perturbations.

It is important to emphasize that the effect of the central body shadow must be included during the evaluation of radiation pressure.
The shadowing routine evaluates, at each time step, whether a particle is occulted by the small body and therefore receives no direct solar illumination. 
This determination is required to apply solar radiation pressure consistently, since the force is switched off when the line of sight to the Sun is blocked. 
The details of the implementation are given in \citet{Regaly2025A&A...697A.116R}.

\subsection{Investigated Ring Systems}
\label{sec:inv-ring-sys}

We investigated the four currently known ring-bearing small body systems associated with (2060) Chiron, (10199) Chariklo, (50000) Quaoar, and (136108) Haumea. 
In all simulations, each central body was propagated on its heliocentric orbit using its observed orbital elements, and the ring was initially constructed as a force-consistent, osculating circular configuration around the body. 

For each system, we performed four models. 
The baseline division was defined by the assumed shape of the central body: in one set, the body was approximated as spherical, while in the other, it was represented by its triaxial figure. 
The spherical models were included to isolate the role of solar radiation pressure by suppressing perturbations arising from the non-spherical gravitational potential. 
Within each of these two baseline cases, we considered two ring-plane configurations: one in which the ring plane was coplanar with the orbital plane of the central body, and another in which the ring plane was inclined by the observed tilt angle.
The ring tilt angle is calculate as $\Delta i=\cos^{-1}(\hat{\mathbf{s}}\cdot\hat{\mathbf{h}})$, where $\hat{\mathbf{s}}$ and $\hat{\mathbf{h}}$ are the ring-pole unit vector and the heliocentric orbital angular-momentum unit vector, respectively.
Thus, a total of four models were computed for each small body system. 

Several small bodies in the Solar System possess multiple rings; however, we modeled only the innermost rings, for which the effect of the triaxial figure is strongest.
The rings were initially assumed to be circular and located at the observed orbital distance from the central body. 
We adopt the same initially narrow ring for all systems as a controlled reference state, in order to isolate radial spreading and vertical thickening driven by solar radiation pressure and triaxial gravity. The initial width of $W_\mathrm{r}=7$~km is a representative narrow-ring scale (comparable to Chariklo’s C1R), while the initial vertical thickness is set to zero because it is poorly constrained observationally. In this setup, the relevant outcome is the increase in ring width and thickness generated by the dynamics, rather than the absolute final dimensions. Our results therefore compare the efficiency of dynamical spreading among systems.

A total of $10^5$ particles were distributed into 10 logarithmically spaced bins characterized by $\beta=[0,~1\times10^{-1},~5\times10^{-2},~2.5\times10^{-2},~1.25\times10^{-2},~6.25\times10^{-3},~3.125\times10^{-3},~1.5625\times10^{-3},~7.8125\times10^{-4},~3.906250\times10^{-4}]$.
We distinguish pebbles from dust by the particles' radiation-pressure parameter $\beta$: pebbles have 
$\beta=0$, whereas dusty particles have $\beta>0$.

Assuming spherical particles composed of water ice, the corresponding particle radii span approximately $s_\mathrm{H_2O}\simeq[0.1~\mathrm{\mu m}-2~\mathrm{mm}]$ based on the optical constants reported by \citet{Warren2008JGRD..11314220W}.
For organic refractory material, a very similar size range is obtained when adopting the optical properties of \cite{Zubko1996MNRAS.282.1321Z}.
In contrast, for metallic iron composition, the same range of 
$\beta$ values corresponds to particle sizes of $s_\mathrm{Fe}\simeq[0.02~\mathrm{\mu m}-2~\mathrm{mm}]$ reflecting the higher material density and different optical response \citep{Henning1996A&A...311..291H}.
A comprehensive summary of the dependence of $\beta$ on particle size for the considered compositions is provided in \citet{Regaly2025A&A...697A.116R}.
The parameters adopted for the investigated systems are summarized in Table~\ref{tab:systems}.

The rings are treated as ensembles of collisionless test particles, such that mutual gravitational interactions are neglected. 
In particular, the self-gravity of the ring is not included in the present model.

The duration of all numerical simulations was 1000~yr.
We used a variable shared time step for the integration, following the simplest scheme based on the fourth-order predictor error given by Eq.~(18) in \citet{Nitadori2008NewA...13..498N}.
To maintain numerical accuracy, the maximum allowed time step, $\Delta t_\mathrm{max}$, was limited to approximately $1/500$ of the orbital period at the ring location.
As a result, we adopted $\Delta t_\mathrm{max}=3$~min for Chariklo, Chiron, and Quaoar, and $\Delta t _\mathrm{max}=1.5$~min for Haumea.

\section{Results}

\begin{figure*}[!h]
    \centering
    \includegraphics[width=0.9\linewidth]{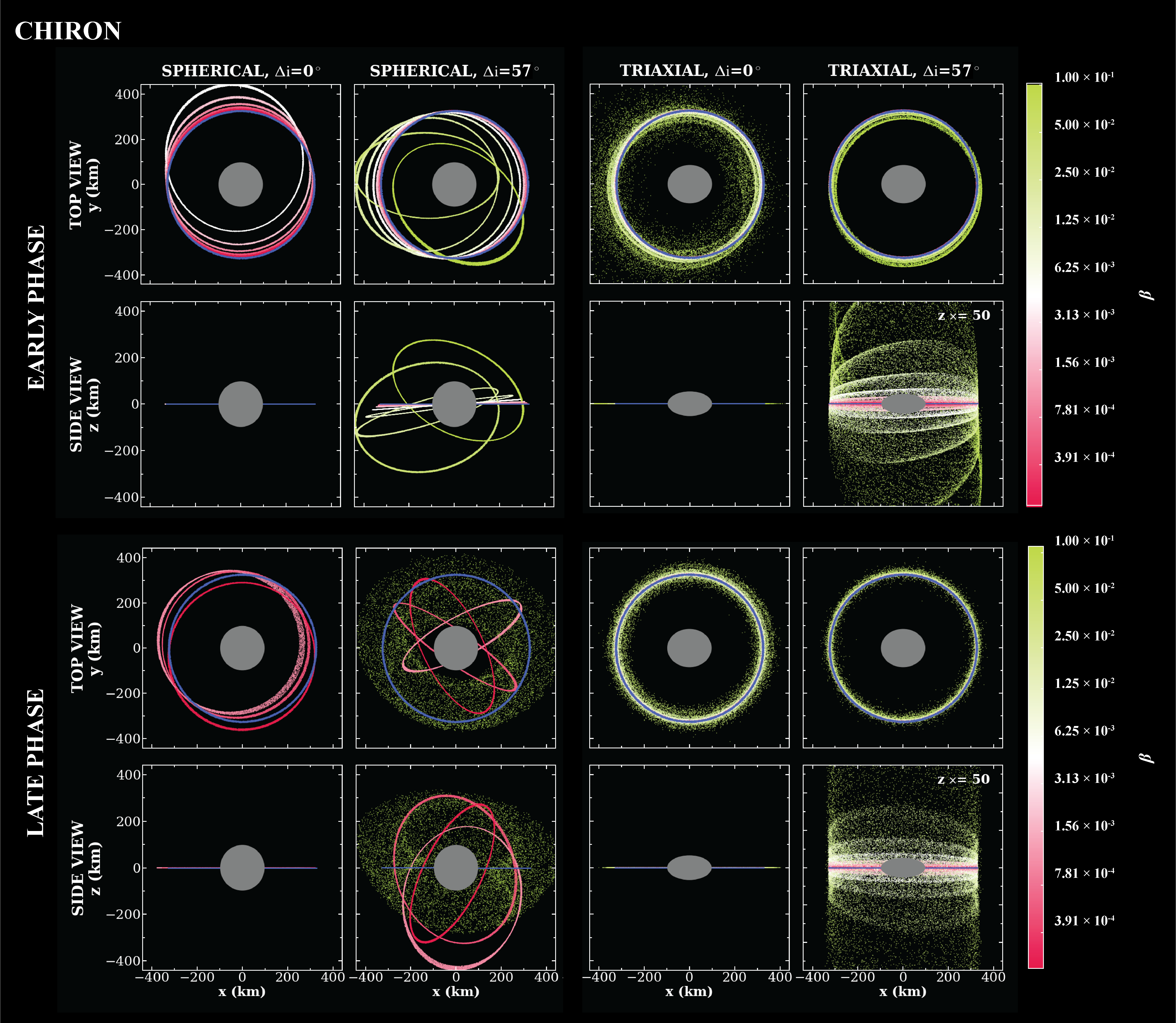}
    \includegraphics[width=0.9\linewidth]{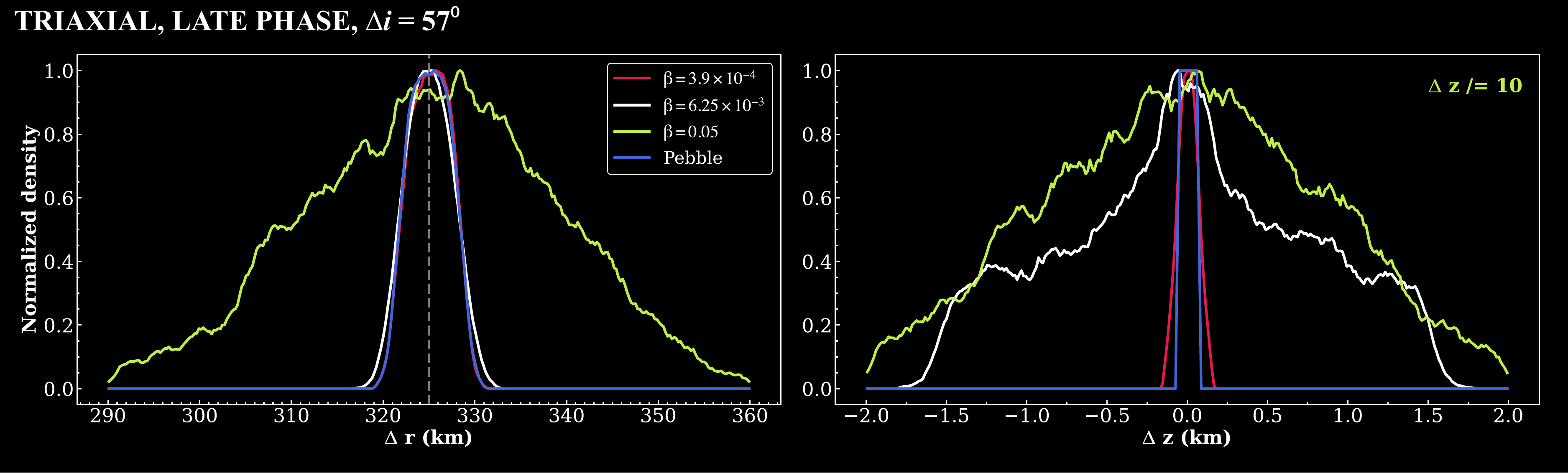}
    \caption{Upper subfigure shows snapshots of ring evolution for the spherical and triaxial Chiron models, shown in top (xy) and side (xz) views. Results are presented for two ring-plane tilt angles, $\Delta i =0^\circ$  and $\Delta i =57^\circ$. The upper panels correspond to an early evolutionary stage ($t\simeq1~$~yr), while the lower panels show a later stage ($t\simeq1000$~yr). Particle sizes, represented by $\beta$, are color-coded from red to lime, while pebbles are represented in blue. 
    For the triaxial models with $\Delta i =57^\circ$, the vertical scale of the ring particle distribution has been artificially exaggerated (by 50 times) in order to emphasize the ring’s vertical expansion.
    Bottom subfigure shows the normalized radial (left panel) and vertical (right panel) density distributions at the late phase. Four particle sizes are distinguished by color, as indicated in the legend: a pebble population ($\beta = 0$) and dust particles with $\beta = 3.9 \times 10^{-4}$, $6.25 \times 10^{-3}$, and $5 \times 10^{-2}$.
    }
    \label{fig:chiron-16}
\end{figure*}

\begin{figure*}
    \centering
    \includegraphics[width=0.9\linewidth]{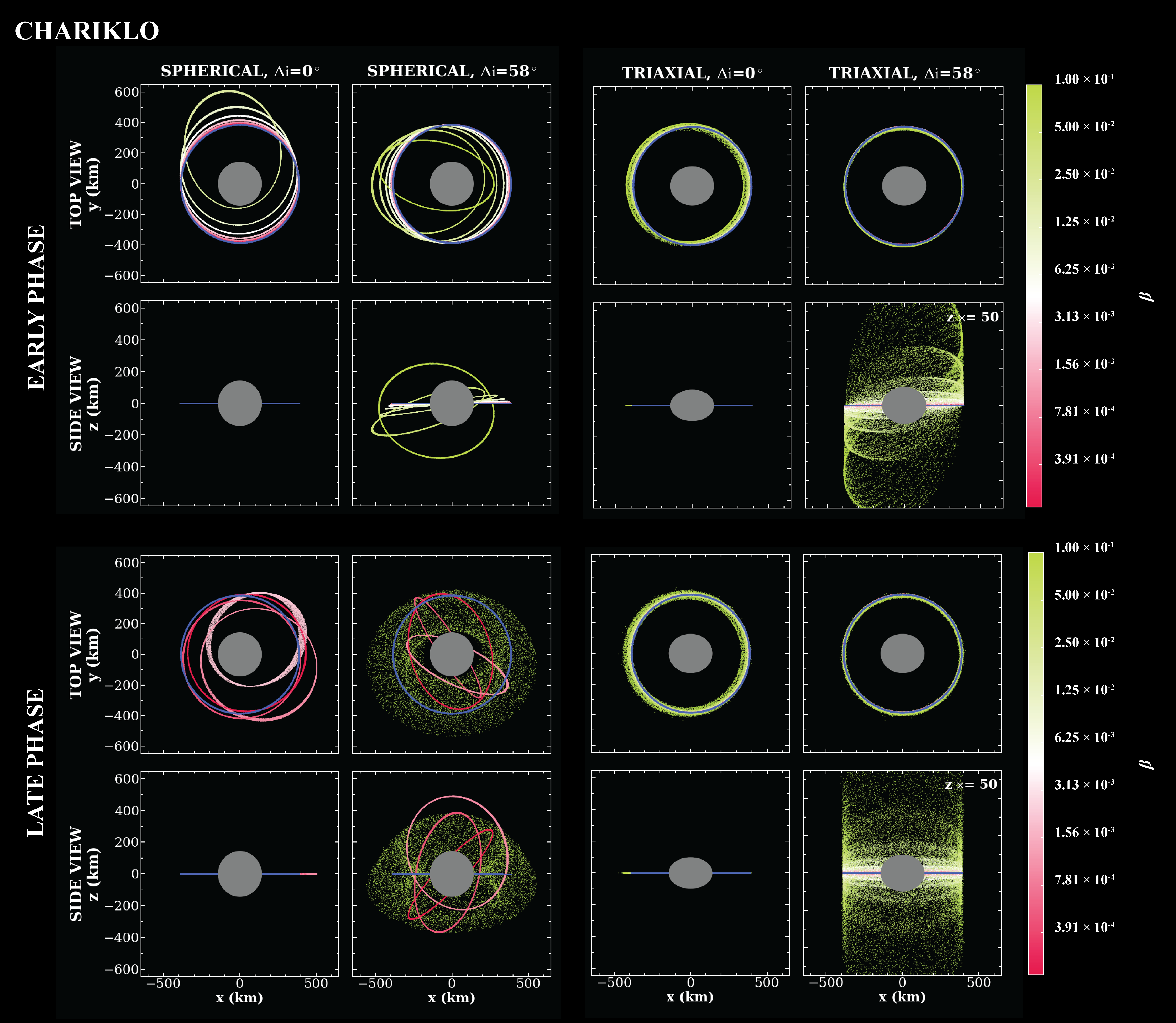}
    \includegraphics[width=0.9\linewidth]{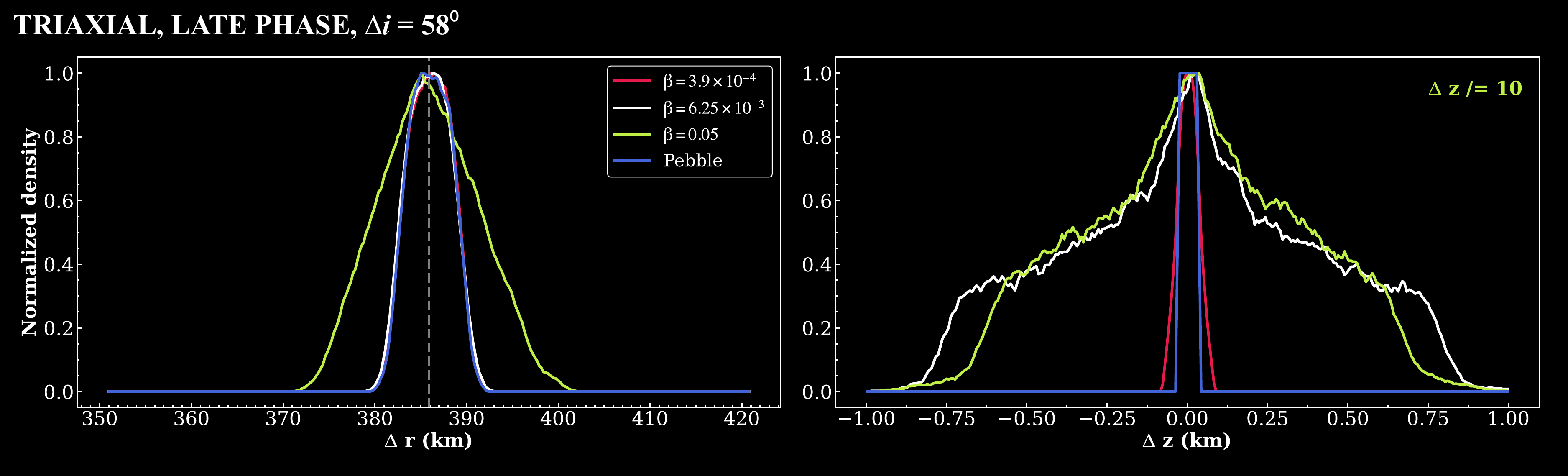}
    \caption{Same as Fig.~\ref{fig:chiron-16} for Chariklo. }
    \label{fig:chariklo-16}
\end{figure*}

\begin{figure*}
    \centering
    \includegraphics[width=0.9\linewidth]{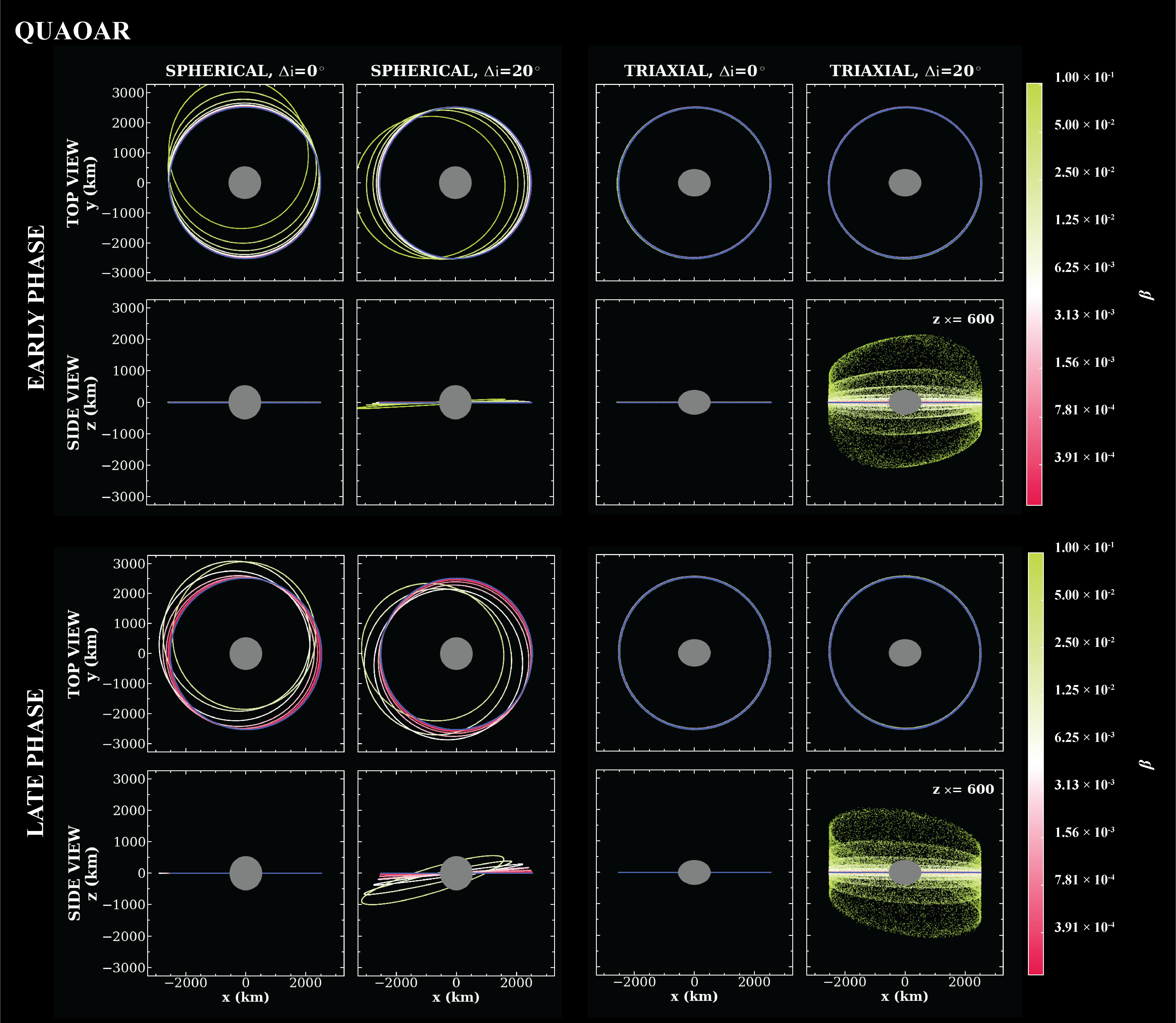}
    \includegraphics[width=0.9\linewidth]{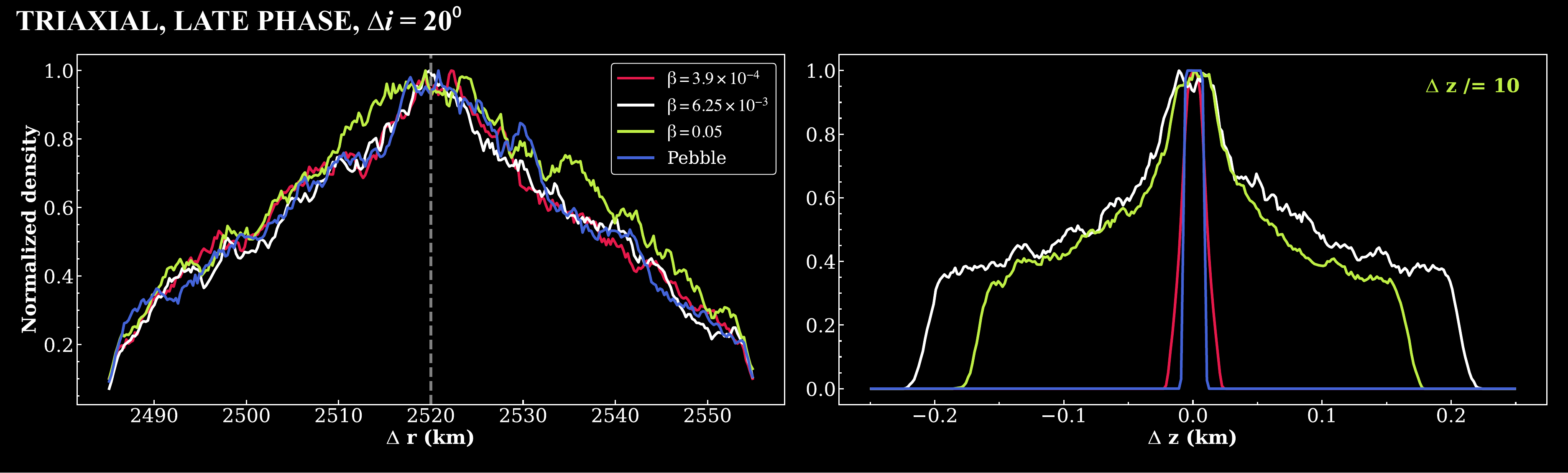}
    \caption{Same as Fig.~\ref{fig:chiron-16} for Quaoar. }
    \label{fig:quaoar-16}
\end{figure*}

\begin{figure*}
    \centering
    \includegraphics[width=0.9\linewidth]{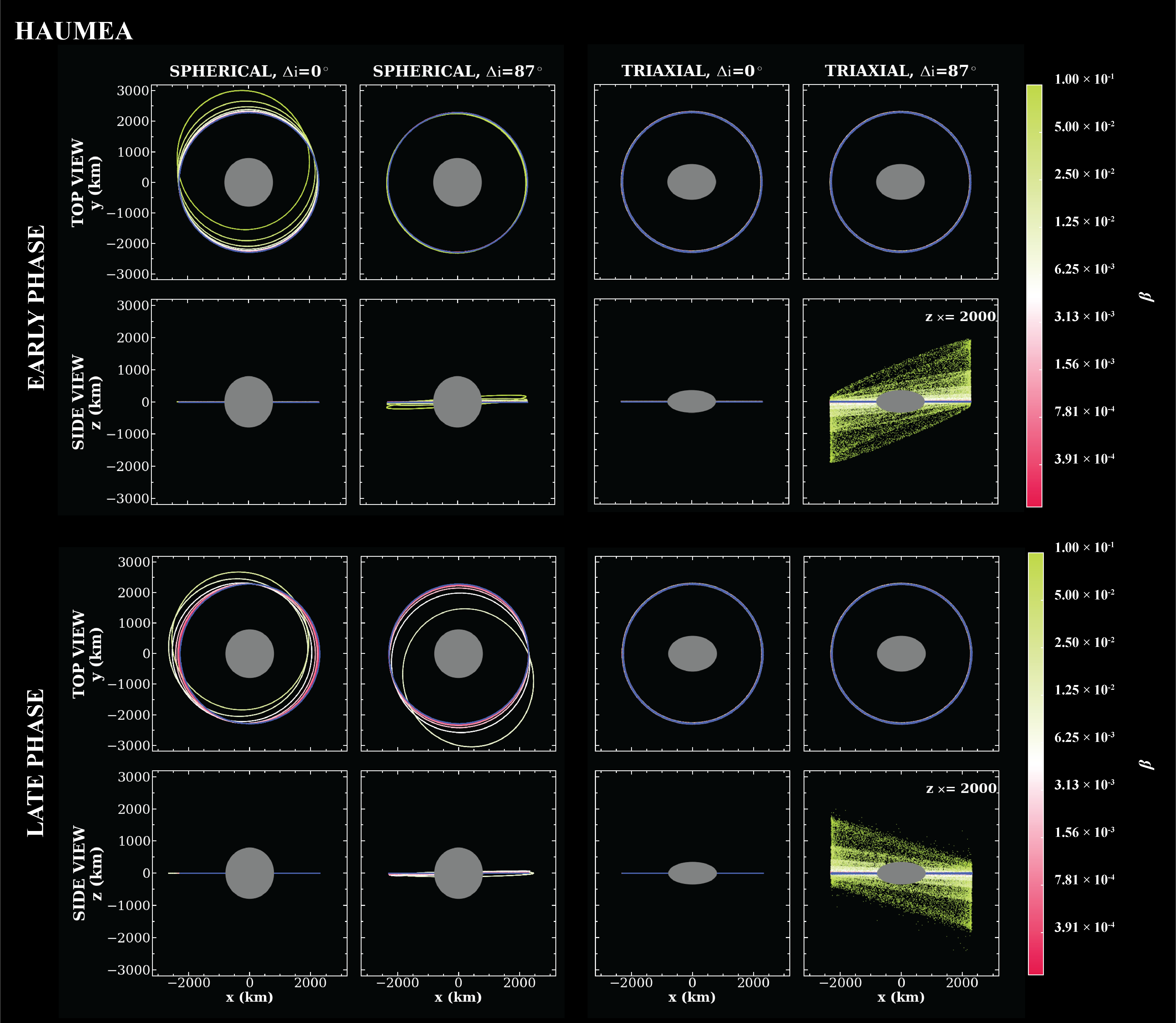}
    \includegraphics[width=0.9\linewidth]{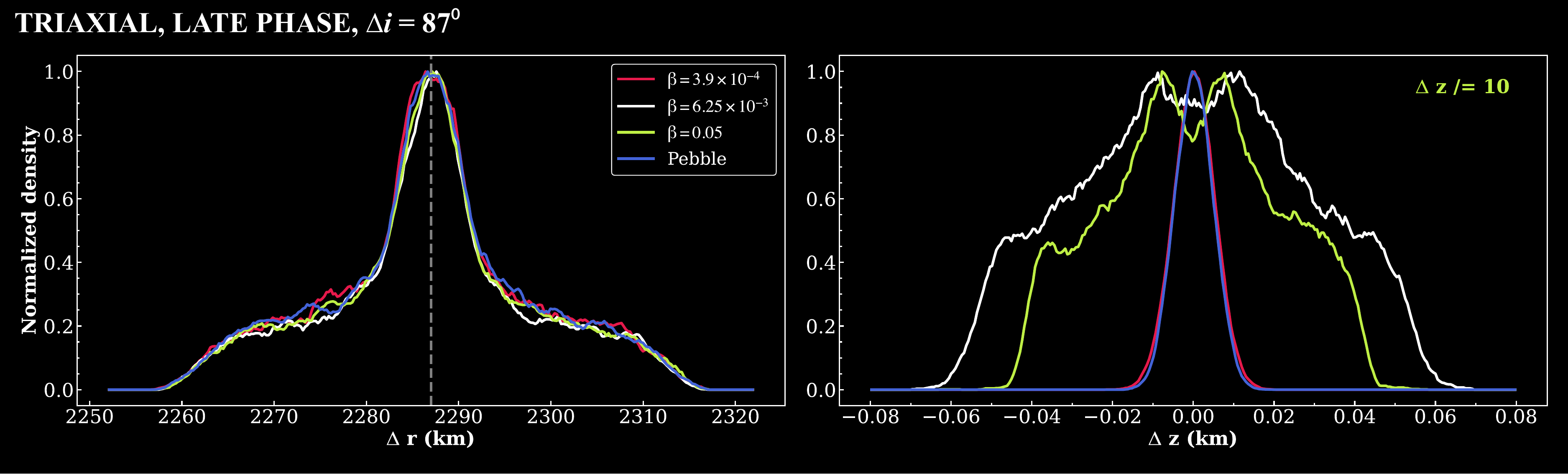}
    \caption{Same as Fig.~\ref{fig:chiron-16} for Haumea. 
    }
    \label{fig:haumea-16}
\end{figure*}

Figures~\ref{fig:chiron-16}--\ref{fig:haumea-16} present representative snapshots of the ring evolution for the four ring-bearing small-body systems considered in this study. 
In each figure, two evolutionary stages are displayed: an early phase ($t=0.1$–$10$~yr, depending on the system) and a late phase ($t=1000$~yr). 
The early phase was set to 0.1~yr for Chiron and Chariklo and 10~yr for Quaoar and Haumea, such that no ring loss occurs in the spherical models and the selected times serve only as illustrative snapshots.
This comparison illustrates both the rapid dynamical depletion of particles assuming a spherical central body with large values of $\beta$, which quickly acquire large eccentricities and are consequently removed from the system, and the longer-term persistence of particles with small $\beta$.
Additional panels (bottom) in these figures show the radial and vertical density profiles of the modeled rings at the late phase for the four bodies in the triaxial cases, assuming that the ring plane is tilted by $\Delta i$ relative to the orbital plane, as specified in Table~\ref{tab:ringed_bodies}.

Several general trends are common to all models in which the central body is treated as spherical.
I.) for pebble-like particles with $\beta\simeq0$, solar radiation pressure is ineffective, and the ring therefore preserves its initial orbital elements over time.
II.) solar radiation pressure excites the eccentricities of the dust ring particles. When the forced eccentricity grows to a critical level such that the particle pericenter becomes comparable to the equivalent radius of the central body, particles are eventually removed through accretion onto the body (see the detailed discussion in \citealp{Regaly2025A&A...697A.116R}). 
As a consequence, the ring lifetime decreases with increasing $\beta$, i.e., with decreasing particle size.
III.) in the inclined-ring configurations, a systematic change of the ring-plane orientation is observed, which behavior should not be confused with the sunflower effect, see in the following.
This behavior is driven by the heliocentric motion of the central body, which modifies the direction of the radiation-pressure forcing over the course of its orbit.
IV.) for inclined rings and the largest radiation-pressure parameter considered ($\beta=0.1$), the particle distribution becomes highly dispersed, and the ring is no longer confined either radially or vertically.
As a result, no coherent ring structure is preserved in this regime, and the radial and vertical density profiles of particles (bottom panels in Figs.~\ref{fig:chiron-16}-\ref{fig:haumea-16}) with $\beta=0.1$ are therefore not presented.

A different set of general behaviors emerges in the triaxial-body models. 
In these cases, the rings do not reach eccentricities large enough for particles to be accreted onto the central body for all investigated $\beta$.
Moreover, for initially inclined rings, the sunflower effect identified by \citet{Regaly2025A&A...697A.116R} is strongly suppressed when the gravitational potential of the central body is triaxial.
The sunflower effect would mean that a highly inclined ring evolves toward a mean plane nearly perpendicular to the incident solar radiation, causing the ring plane to rotate with the host body’s heliocentric motion and remain approximately Sun-facing.
Rather than producing a pronounced systematic reorientation of the ring plane, the dominant outcome is an increase in the mean vertical thickness of the ring. 
This vertical broadening is directly proportional to $\beta$, implying that particles with larger $\beta$ values develop a larger vertical extent.  A prime example is the Phoebe ring around Saturn; however, in this case, other effects also contribute to the broadening \citep{Tamayo2016}.

We now compare the individual models. 
Two categories can be identified: systems associated with lower-mass central bodies (Chiron and Chariklo) and those with higher-mass bodies (Quaoar and Haumea).
In the spherical models of Chiron (Fig.~\ref{fig:chiron-16}) and Chariklo (Fig.~\ref{fig:chariklo-16}), particles with $\beta \gtrsim 3\times10^{-3}$ are lost from the system, independent of whether the ring is initially located in the orbital plane or in an inclined configuration. 
In the inclined spherical models, a well-defined ring plane cannot be maintained; instead, the apparent ring-plane orientation becomes dependent on $\beta$.
It is important to note that the orientation of these particle distributions is not stationary: the effective ring planes change continuously as the central body progresses along its heliocentric orbit.

In the triaxial models of Chiron and Chariklo, all particles remain confined to the ring regardless of the value of $\beta$. 
However, particles with $\beta=0.1$ undergo substantial radial and vertical spreading, and their distributions are therefore omitted.
As shown in the lower-left panels of Figs.~\ref{fig:chiron-16} and \ref{fig:chariklo-16}, the radial ring width, $W_\mathrm{r}$ (defined as the width over which the normalized density exceeds 20 percent), reaches a relatively large values of $W_\mathrm{r}\simeq70$~km for Chiron and $W_\mathrm{r}\simeq30$~km for Chariklo.
However, for $\beta \lesssim 6.25\times10^{-3}$ the 
radial width becomes nearly independent of $\beta$ and remains comparable to the width of the pebble ring, 
$W_\mathrm{r}\simeq10$\,km in both cases.
The dust rings develop vertical scale heights, which depend sensitively on $\beta$ (see, lower-right panels of Figs.~\ref{fig:chiron-16} and \ref{fig:chariklo-16}). 
For Chiron and Chariklo, the vertical thickness reaches approximately $W_\mathrm{z}\simeq40$~km and $W_\mathrm{z}\simeq15$~km, respectively, for particles with $\beta=5\times10^{-2}$. 
Note that, for clarity, the vertical scale height displayed for $\beta=5\times10^{-2}$ has been reduced by a factor of 10. However, at the critical value $\beta=6.25\times10^{-3}$, the vertical ring thickness decreases to $W_\mathrm{z}\simeq1.75$~km for Chiron and $W_\mathrm{z}\simeq0.4$~km for Chariklo.

We now turn to the more massive systems, Quaoar and Haumea (see Figs.~\ref{fig:quaoar-16} and \ref{fig:haumea-16}, respectively). 
In the spherical models, the central bodies are sufficiently massive to retain dust ring particles with $\beta \lesssim 5\times10^{-2}$. 
However, for $\beta \gtrsim 6.25\times10^{-3}$, the rings develop significant eccentricities, with eccentricity increasing approximately in proportion to $\beta$. 
By contrast, in the triaxial models, the ring eccentricity remains small, typically of order $10^{-3}$--$10^{-2}$, depending on $\beta$, the magnitude of $C_\mathrm{22}$, and the mass of the central body. 
As indicated by the radial density profiles (see lower-left panels of Figs.~\ref{fig:quaoar-16} and \ref{fig:haumea-16}), the radial width of the ring in these models is nearly independent of $\beta$. 
The rings of Quaoar and Haumea exhibit substantial radial widths, with $W_\mathrm{r}\simeq40$~km and $W_\mathrm{r}\simeq70$~km, respectively, owing both to eccentricity excitation by their massive, elongated central bodies and to ring radii that are an order of magnitude larger than those around less massive bodies modeled.
Concerning the vertical scale height of the rings (see the lower-right panels of Figs.~\ref{fig:quaoar-16} and \ref{fig:haumea-16}), we find the same qualitative behavior as in the less massive systems; however, in these cases the vertical scale heights are smaller by approximately one to two orders of magnitude. For $\beta=6.25\times10^{-3}$, we obtain $W_\mathrm{z}\simeq0.4$~km for Quaoar and $W_\mathrm{z}\simeq0.12$~km for Haumea. 
The relatively small vertical thickness of the ring is a consequence of a purely gravitational effect: the central body -- three orders of magnitude more massive and significantly oblate -- suppresses large inclination excitation.

\section{Discussion}
\label{sec:discussion}

\subsection{The Stabilizing Role of Oblateness}

For small eccentricities, it is convenient to use the complex eccentricity $Z\equiv e \exp(i\varpi)$, where $e$ is the eccentricity and $\varpi$ is the longitude of pericenter. 
A standard orbit-averaged description of weak, slowly varying, in-plane perturbations yields a forced response whose amplitude is controlled by the mismatch between the forcing pattern speed
and the particle's apsidal precession rate (e.g. \citealt{HamiltonKrivov1996,Hedman2010,Regaly2025A&A...697A.116R}).
In particular, for RP forcing with an inertial direction that circulates at the heliocentric mean motion $n_\odot$, the forced eccentricity scales as
\begin{equation}\label{eq:ef_scaling}
e_{\rm f}
\;\simeq\;
\frac{a_{\rm RP}}{n a~\left|\dot{\varpi}-n_\odot\right|}
\;=\;
\left(\frac{a_{\rm RP}}{n^{2}a}\right)~
\frac{n}{\left|\dot{\varpi}-n_\odot\right|}~ ,
\end{equation}
where at a body's heliocentric distance, the direct solar radiation-pressure acceleration has a magnitude of $a_\mathrm{RP}=|\boldsymbol a_{\rm RP}|$ (see, Eq.~\ref{eq:arp}).
Thus, the key control parameter is the apsidal precession rate $\dot{\varpi}$.

For a strictly spherical central potential, nearly Keplerian orbits satisfy (to lowest order) $\dot{\varpi}\approx 0$, so the denominator in Eq.~(\ref{eq:ef_scaling}) can become small because the RP forcing direction changes only slowly ($n_\odot$ is small already at $\sim$15--20 AU). 
This makes the RP forcing coherent over many body-centric orbits, allowing the eccentricity vector to grow to large amplitudes
(or to reach a large forced equilibrium), consistent with classical circumplanetary dust dynamics \citep{Burns1979Icar...40....1B,HamiltonKrivov1996}.

Once $e$ becomes sufficiently large, the particle's pericenter distance $q = a(1-e)$ can fall below the physical radius $R$ of the small body, and engulfment occurs for $e> 1-(R/a)$.
Therefore, in a spherical model (small $\dot{\varpi}$), RP can drive $e$ upward until sufficiently small particles (large $\beta$) produce an accretion channel \citep{Burns1979Icar...40....1B,Regaly2025A&A...697A.116R}.

However, as shown in the previous section, oblateness suppresses RP-driven eccentricity.
If the small body is axisymmetric but oblate, its dominant zonal harmonic $J_2$ induces apsidal precession. For low inclination orbits (appropriate for equatorial rings),
\begin{equation}\label{eq:j2_precession}
\dot{\varpi}_{J_2}
\;\simeq\;
\frac{3}{2}~J_2~n\left(\frac{R}{a}\right)^2
\quad (i\approx 0)~ .
\end{equation}
As such, even a modest $J_2$ can make $|\dot{\varpi}-n_\odot|$ large compared to the spherical case.
Equation~\eqref{eq:ef_scaling} then implies
\begin{equation}
e_{\rm f}\;\propto\;\frac{1}{|\dot{\varpi}-n_\odot|}~ ,
\end{equation}
so an oblate figure reduces the forced eccentricity and prevents the large radial excursions
that lead to impact. 
This is precisely the mechanism invoked, for example, in Saturn’s dusty rings (the E ring) and in dust associated with Phobos, as well as in the so-called Charming Ringlet, where strong quadrupolar precession detunes RP forcing and limits the growth of 
eccentricity (\citealt{HamiltonKrivov1996,Hedman2010}).

\subsection{$C_{22}$-driven Resonance Excitation}

A triaxial figure introduces the tesseral harmonic $C_{22}$ (Eq.~\ref{eq:U_J2C22}).
The essential change relative to the purely oblate case is that the $C_{22}$ term is non-axisymmetric: it applies an azimuthal force component that can exchange angular momentum with the orbit in the inertial frame. 
Consequently: I.) the semi–major axis can undergo long-period variations (not purely secularly constant as in a strictly axisymmetric field), II.) the dynamics in the rotating frame admits corotation points whose stability depends on $C_{22}$ and rotation rate, III.) and, most importantly for rings, SORs appear.
As a result, the rotating triaxial potential provides an additional mechanism for eccentricity excitation, thereby contributing to ring broadening.

\begin{figure}
    \centering
    \includegraphics[width=1\linewidth]{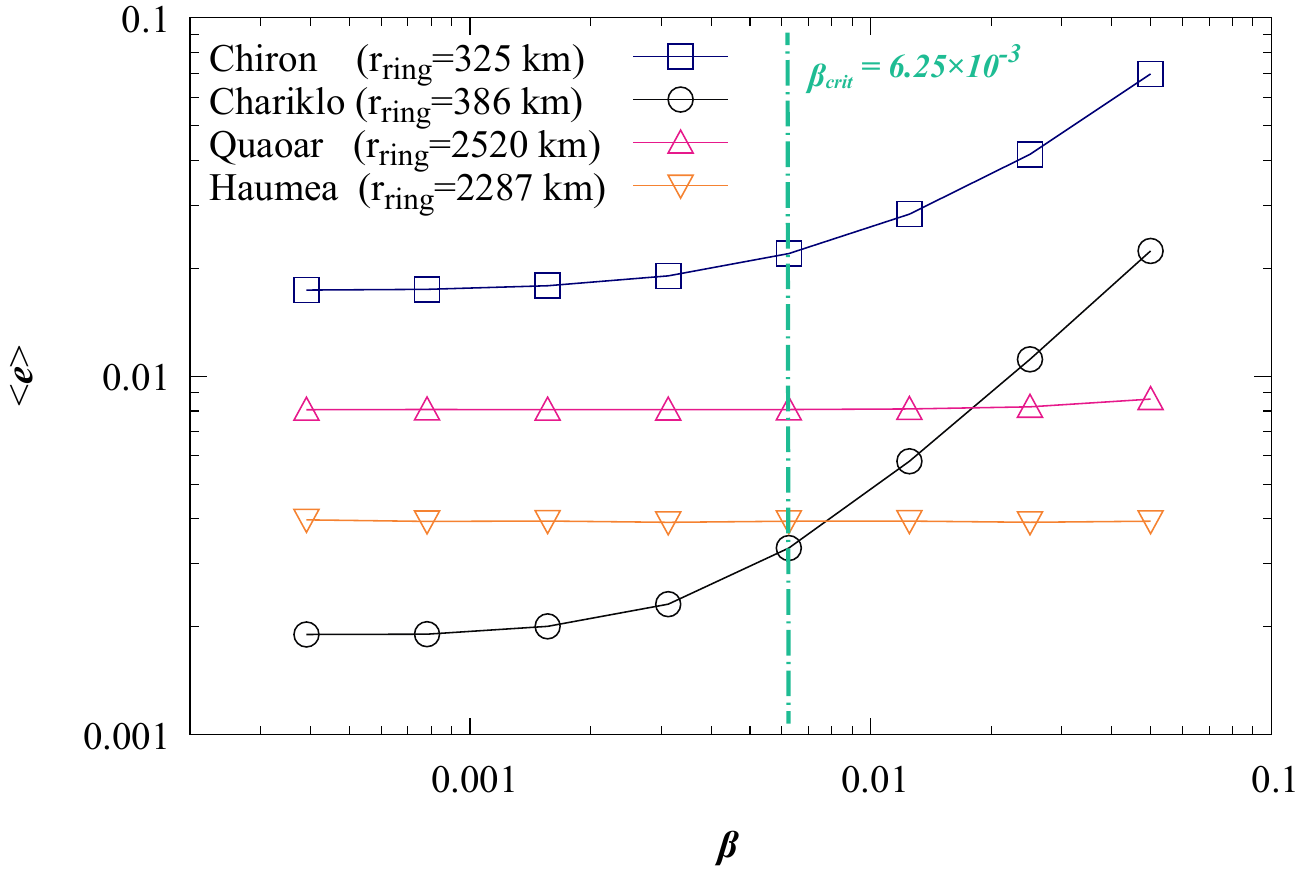}
    \caption{Average eccentricity of ring particles, $\left< e \right>$, as a function of $\beta$, measured for the four triaxial central-body models.}
    \label{fig:ringecc}
\end{figure}

From radial fluctuations of ring particles with a given $\beta$ located at a distance $r_{\mathrm{i},\beta}(t)$ from the central body, the mean eccentricity can be approximated as
\begin{equation}
\langle e \rangle\simeq\sqrt{2}\,\frac{\sigma_{r,\mathrm{i},\beta}}{\langle r_{\mathrm{i},\beta}(t) \rangle},
\end{equation}
where $\sigma_{r,\mathrm{i},\beta}$ denotes the standard deviation of $r_{\mathrm{i},\beta}$ about $\langle r_{\mathrm{i},\beta}(t) \rangle$ \citep{MurrayDermott1999ssd..book.....M}. 
As shown in Fig.~\ref{fig:ringecc}, $\langle e \rangle$ lies in the range $2\times10^{-3}$ to $2\times10^{-2}$ for $\beta \lesssim 6.25\times10^{-3}$. 
Notably, for Quaoar and Haumea, $\langle e \rangle$ is largely insensitive to $\beta$, indicating that the measured eccentricity is primarily generated by the elongation of the triaxial figure represented by $C_{22}$ rather than RP.
Note that the eccentricity excitation driven by radiation pressure depends not only on $\beta$, but also on the strength and geometry of the gravitational potential.
In contrast, for Chiron and Chariklo, once $\beta$ exceeds the critical value of $6.25\times10^{-3}$, radiation pressure produces a substantially larger eccentricity.

At the lowest values of the radiation parameter $\beta$, the direct effect of RP is negligible and the measured eccentricity is controlled primarily by the rotating non-axisymmetric gravitational field of the central body. 
In that limit, the particle dynamics are governed mainly by the triaxial component of the potential (through the $C_{22}$ term). 
The resulting eccentricity excitation depends on three coupled factors: the strength of the effective triaxial forcing, the ring distance, and the proximity of the adopted ring radius to the central body.
Our simulations show a clear classification in the forced eccentricity at low $\beta$: Chiron exhibits the largest eccentricity, Chariklo the smallest, and Quaoar a larger response than Haumea. 
This ordering cannot be understood from triaxiality alone; rather, it reflects how the adopted ring radii sample different parts of the phase space associated with the rotating figure.

For Chiron and Chariklo, ring locations are close to the  $1:3$ SOR, but their dynamical responses are markedly different. 
A natural interpretation is that Chiron experiences a stronger effective non-axisymmetric forcing at its adopted ring distance, either because of a larger effective $C_{22}$ term, a more favorable geometrical scaling of the perturbation, or a closer approach to a SOR. 
In contrast, the ring of Chariklo appears to lie in a comparatively quiet region of phase space, where the triaxial forcing is weak, and the associated forced eccentricity remains very small. 

The contrast between Quaoar and Haumea is also instructive. 
Although Haumea is the more elongated body, its ring does not necessarily probe the most strongly excited part of the $1:3$ resonant structure. 
Previous dynamical studies have shown that the exact $1:3$ resonant family around Haumea is accompanied by a separatrix and a chaotic layer, and that the corresponding resonant periodic orbits require relatively large radial excursions \citep{Ribeiro2023}.
This makes the observed ring location more naturally associated with a low-eccentricity family of non-resonant or weakly resonant stable trajectories rather than with the exact resonant core. 
Consequently, a particle ensemble started at the observed Haumea ring distance can display only moderate mean eccentricities despite the strong triaxiality of the central body.

By contrast, the larger forced eccentricity found at Quaoar can be interpreted as evidence that the adopted ring radius lies closer to an SOR. 
Since the Quaoar ring has been discussed in connection with the $5:7$ SOR \citep{Pereira2023QuaoarRings,Sicardy2025}, the larger eccentricity in the simulations is consistent with the ring sampling a more effective forcing region than in the Haumea case. 
Thus, the Quaoar--Haumea comparison indicates that the observed eccentricity response is controlled less by the absolute figure elongation of the body than by the detailed phase-space location of the ring relative to the relevant resonant structure.

Overall, the low-$\beta$ results suggest the following physical picture. 
Chiron corresponds to the strongest effective triaxial forcing among the four systems considered here; Chariklo occupies the most weakly forced configuration; Haumea lies near a dynamically complex but comparatively low-eccentricity stable branch; and Quaoar appears to be located closer to an eccentricity-exciting commensurability.

\begin{figure}
    \centering
    \includegraphics[width=1\linewidth]{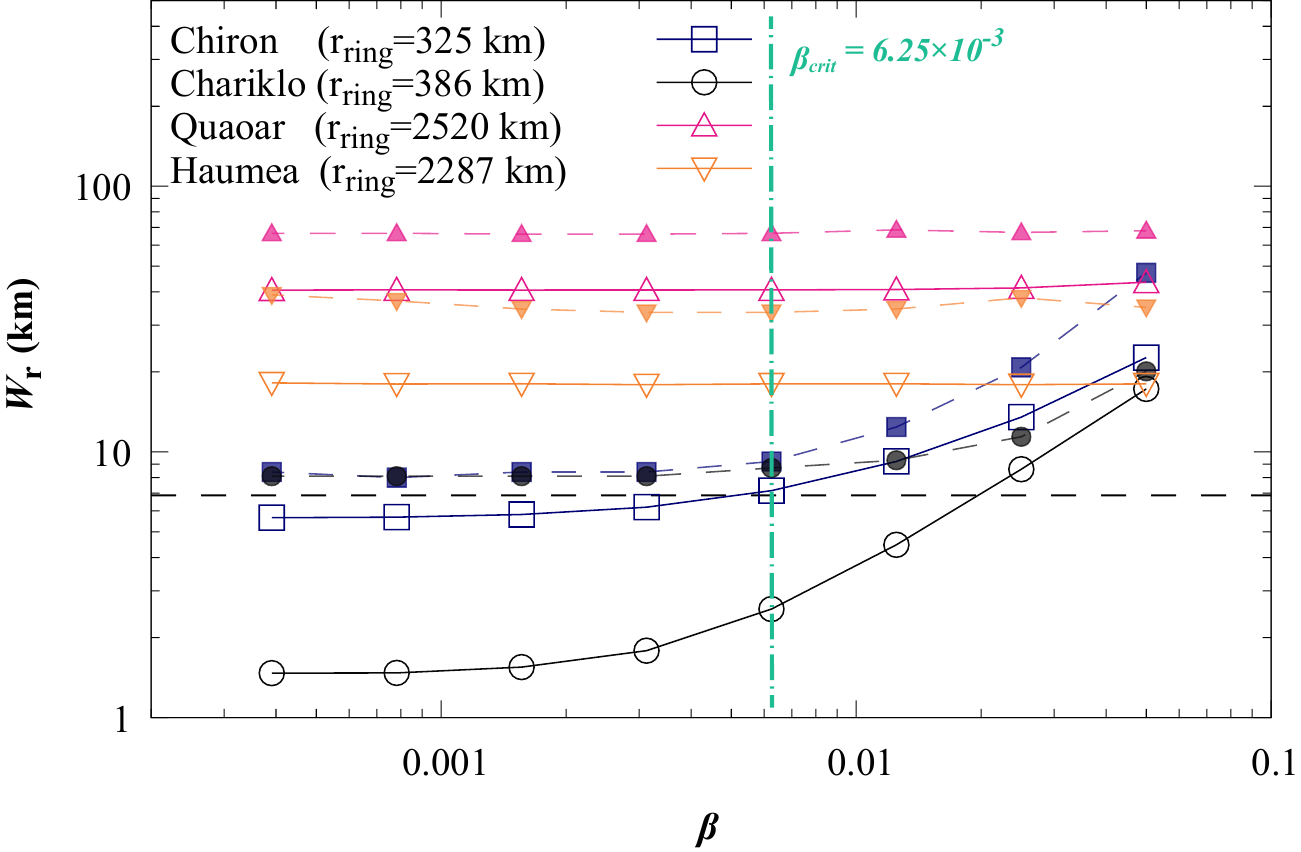}
    \caption{The ring radial width, $W_\mathrm{r}$, determined from the numerical simulations above 20 percent normalized density (shown with filled symbols).
    The radial spread associated with eccentric particle orbits, $\delta r_\mathrm{ring}$, inferred from the mean eccentricity excitation of ring particles for a given $\beta$ (shown with open symbols).  
    The dashed black lines mark the initial ring width.}
    \label{fig:ringwidth}
\end{figure}

Since the ring width reflects radial spreading, it is expected to correlate with $\langle e \rangle$.
The spreading is driven by the forced eccentricity and is given by
\begin{equation}
\delta r_\mathrm{ring} = r_\mathrm{ring}\left[(1+\langle e\rangle)-(1-\langle e\rangle)\right]
= 2r_\mathrm{ring}\langle e\rangle.
\end{equation}
Figure~\ref{fig:ringwidth} compares the measured values of $W_\mathrm{r}$ (defined as the width over which the normalized density exceeds 20 percent) obtained from the simulations with the corresponding values of $\delta r_\mathrm{ring}$ as a function of $\beta$.
Owing to the finite initial width of the ring, $W_\mathrm{r} > \delta r_\mathrm{ring}$.
It is noteworthy that, for $\beta \lesssim 6.25\times10^{-3}$, $W_\mathrm{r} \lesssim 10$~km for Chiron and Chariklo; however, the rings of Quaoar and Haumea spread much more significantly. This difference is driven by the forced eccentricity and by the roughly ten times larger ring radii in these systems.

The measured ring widths for Chariklo, Chiron (assuming $\beta\lesssim6.25\times10^{-3}$) and Haumea are consistent with observational estimates. 
However, they are substantially larger than the width measured for Quaoar’s innermost ring, which is approximately $10$~km \citep{Pereira2023QuaoarRings}.
We speculate that this discrepancy may be explained either by Quaoar having a shorter rotational period -- perhaps by a factor of two, as may be plausible in the tidal evolution models of \citealp{Regaly2025PASP..137k4401R} -- or by Quaoar having a more spherical shape. 
A detailed investigation of these possibilities is postponed to future work.

\subsection{Caveats}

Here, we briefly summarize the main limitations of our simulations.
First, we neglect inter-particle collisions and the self-gravity of the ring. 
In dense rings, self-gravity and collisions collectively act as an effective viscosity, redistributing angular momentum and damping random motions. 
Such collective effects tend to suppress coherent eccentricity growth and can modify both the secular evolution and the stability boundaries of the ring \citep[see e.g.][for the case of Chariklo]{Michikoshi2017Chariklo}.

Consistent with this expectation, faster apsidal precession enhances phase mixing of the RP-driven eccentricity vector within the ring. 
In a collisional environment, this phase dispersion facilitates the damping of coherent eccentricity through inelastic collisions and collective gravitational interactions before the impact threshold is reached \citep{HamiltonKrivov1996}. 
Therefore, our collisionless treatment likely overestimates the maximum eccentricities attainable under radiation-pressure forcing.

Moreover, in collisional rings, SORs associated with the $C_{22}$ harmonic can, in principle, provide torques that partially counteract viscous spreading. However, these same resonances also tend to excite eccentricities and may induce streamline crossing when the forced $e$ becomes sufficiently large \citep{Sicardy2025}. Compared to the purely oblate case, the inclusion of $C_{22}$ therefore introduces narrow semimajor-axis intervals where the dynamics is resonance-dominated, bordered by regions that may be more strongly stirred or dynamically chaotic.

We neglect the Poynting–Robertson (PR) drag in our present simulations. 
PR drag removes angular momentum from ring particles at a rate proportional to $\beta$. 
While PR drag ultimately drives orbital decay and leads to the gradual destruction of the ring, its characteristic timescale at heliocentric distances of interest is typically several hundred million years, substantially longer than the duration covered by our integrations \citep{Hyodo2025ApJ...994L..15H}. 
Consequently, PR drag does not affect the short-term dynamical mechanisms explored here, although it will control the long-term survival of the ring. 
A future study will incorporate collisional physics, self-gravity, and PR drag to assess their coupled influence on ring evolution.

We neglect the classical Yarkovsky force and its eclipse-modulated analogue in the present integrations.
The Yarkovsky effect is a secular thermal-recoil force that requires a persistent temperature asymmetry on a rotating body with non-zero thermal inertia, and its standard treatment is primarily developed for coherent meteoroids and asteroid fragments rather than for collisional dust particles in a dense particulate ring \citep{Vokrouhlicky1999,Bottke2006}.
The Yarkovsky effect requires a finite thermal lag, usually parameterized by thermal inertia. 
In a sufficiently small particle, heat is redistributed so quickly that the particle becomes nearly isothermal over the relevant forcing timescale.
Therefore, over the short simulated interval of $ 10^{3}~\mathrm{yr}$, both effects should remain subdominant for the sub-millimeter ring particles considered here.

A similar argument applies to the eclipse-modulated (Yarkovsky--Schach-like) contribution: although shadowing by the central body can, in principle, introduce an additional thermal asymmetry, a measurable secular drift requires repeated phase-coherent eclipses over long intervals \citep{Rubincam1988,Zhou2024}. 
By analogy with binary-asteroid calculations, where such thermally driven orbital evolution typically operates on \(\gtrsim 10^{5}~\mathrm{yr}\) timescales, the corresponding effect in these much more weakly illuminated outer-Solar-System ring systems should be even slower \citep{Zhou2024,Zhou2026arXiv260302585Z}. 
We therefore do not expect either the direct or eclipse-modulated Yarkovsky terms to modify the ring morphology appreciably over \(\sim 10^{3}~\mathrm{yr}\), although they may become relevant in calculations aimed at long-term secular migration.

\section{Conclusion}

In this study, we investigated the influence of solar radiation pressure (RP) on rings composed of sub-millimeter particles around the Solar System small bodies (2060) Chiron, (10199) Chariklo, (50000) Quaoar, and (136108) Haumea. 
We account for the host body’s non-axisymmetric gravity field by including the second-order terms in the gravitational potential (Eq.~\ref{eq:U_J2C22}). 
For triaxial figures, we also include the body’s rotation. 
In addition, we model RP together with eclipses due to the host-body shadow. 
The heliocentric motion of each system is incorporated using the objects’ known orbital elements. 
All simulations are performed with our GPU-accelerated fourth-order Hermite integrator, HIPERION.
Our main conclusions are the following.

1) Assuming a spherical central body, radiation pressure drives eccentricity growth efficiently. 
Above a critical value of $\beta$ ($3\times10^{-3}$ for Chariklo and Chiron, $5\times10^{-2}$ for Quaoar and Haumea), the forced eccentricity exceeds the impact threshold, and the ring is lost rapidly due to RP-driven eccentricity excitation (see details in \citealp{Regaly2025A&A...697A.116R}).

2) By contrast, adopting the realistic rotating triaxial shape of the body changes this outcome. 
The dominant $J_2$ harmonic of the gravitational potential induces sufficiently rapid apsidal precession to de-tune the radiation-pressure forcing. 
As a result, for all considered $\beta$ values (corresponding to sub-millimeter particle sizes), impact-causing eccentricity excitation due to RP remains confined over the simulated interval.

3) Assuming a realistic triaxial shape for the central body, a strongly confined ring structure forms for $\beta \lesssim 6.25\times10^{-3}$ (corresponding to particle sizes $\gtrsim 7$--$40~\mu$m, depending on the material composition; see Sect.~\ref{sec:inv-ring-sys}). 
In the cases of Chariklo and Chiron, such confined rings exhibit radial widths of order $W_\mathrm{r}\simeq10$~km and vertical thicknesses of $W_\mathrm{z}\lesssim0.4$--$1.75$~km for $\beta \lesssim 6.25\times10^{-3}$. 
For larger $\beta$, RP-driven eccentricity excitation exceeds that induced by the triaxial $C_{22}$ term, leading to a substantial radial broadening of the ring. 

4) By contrast, for triaxial model of Quaoar and Haumea, the ring width remains approximately constant at $W_\mathrm{r}\simeq40-70$~km, essentially independent of $\beta$, with vertical thicknesses of $W_\mathrm{z}\lesssim0.12$--$0.4$~km. 
The enhanced radial spreading of the ring, relative to that in the lower-mass counterparts, arises from the combined effects of stronger eccentricity excitation by the large $C_{22}$ component of the triaxial gravitational potential, as indicated by the very weak dependence of the eccentricity excitation on $\beta$ (Fig.\ref{fig:ringecc}), and the approximately ten times larger ring diameters.

5) The RP-driven configuration identified by \citet{Regaly2025A&A...697A.116R}, in which a highly inclined ring around a spherical central body evolves toward a state whose mean plane remains nearly perpendicular to the incident solar radiation -- so that the ring plane co-rotates with the host body’s heliocentric motion and remains approximately Sun-facing -- does not occur in our triaxial models.
In the presence of a triaxial gravitational field, the additional quadrupolar and tesseral harmonics modify the secular precession and suppress the RP-driven alignment mechanism found in the spherical case.

We conclude that, over the simulated time interval, the dynamical evolution of narrow rings around small bodies is fundamentally controlled by the competition between radiation-pressure forcing and apsidal precession induced by the gravitational field.
While spherical models predict efficient eccentricity growth and ring loss above a critical $\beta$, incorporating the realistic triaxial gravity field stabilizes sub-millimeter particles $\gtrsim7-40~\mu$m through rapid precession and resonance structure, leading to confined and dynamically robust ring configurations. 
These results demonstrate that accurate modeling of higher-order gravitational harmonics is essential for assessing the survival and morphology of dusty rings around irregular small bodies.

\section*{acknowledgments}
This research was partially supported by the National Research, Innovation and Development Office (NKFIH), Hungary, through the grant K-138962. 
We thank the anonymous referee for their constructive comments, which significantly improved the quality of the paper.

\bibliography{RingPebbleDust}{}
\bibliographystyle{aasjournal}

\appendix

\section{Triaxial Terms for a 4th-Order Hermite Scheme}
\label{apx:acc_jerk}

In this appendix, we present an efficient method for calculating the potential and its first and second derivatives, following \citet{Cunningham1970CeMec...2..207C}.
Let $\mathbf{r}_\mathrm{b}=(x_\mathrm{b},y_\mathrm{b},z_\mathrm{b})$ denote body-centric Cartesian coordinates aligned with the principal axes (with $z_\mathrm{b}$ along the spin axis).
The potential given in Eq.~\eqref{eq:pot} at a body-centric position $\mathbf{r}_\mathrm{b}$ is 
\begin{equation}
    V_\mathrm{b}(\mathbf{r}_\mathrm{b})=\frac{\mu_\mathrm{b}}{r_\mathrm{b}}\left(\frac{R_\mathrm{eq}}{r_\mathrm{b}}\right)^2\left[\frac{1}{2}J_{2}\left(\frac{3z_\mathrm{b}^2-r_\mathrm{b}^2}{r_\mathrm{b}^2}\right)-3C_{22}\left(\frac{x_\mathrm{b}^2-y_\mathrm{b}^2}{r_\mathrm{b}^2}\right)\right].
    \label{eq:U_J2C22}
\end{equation}
The accelerations caused by the zonal and tesseral harmonic are
\begin{equation}
\mathbf{a}_\mathrm{J2} =
k
\renewcommand{\arraystretch}{1.5}
\begin{bmatrix}
\displaystyle x_\mathrm{b}\left(5\frac{z_\mathrm{b}^2}{r_\mathrm{b}^2}-1\right)\\
\displaystyle y_\mathrm{b}\left(5\frac{z_\mathrm{b}^2}{r_\mathrm{b}^2}-1\right)\\
\displaystyle z_\mathrm{b}\left(5\frac{z_\mathrm{b}^2}{r_\mathrm{b}^2}-3\right),
\end{bmatrix}
\label{eq:aJ2}
\end{equation}
\begin{equation}
\mathbf{a}_\mathrm{C22} =
q
\renewcommand{\arraystretch}{1.5}
\begin{bmatrix}
\displaystyle \frac{2x_\mathrm{b}}{r_\mathrm{b}^5}-\frac{5x_\mathrm{b}D}{r_\mathrm{b}^7}\\[4pt]
\displaystyle -\frac{2y_\mathrm{b}}{r_\mathrm{b}^5}-\frac{5y_\mathrm{b}D}{r_\mathrm{b}^7}\\[4pt]
\displaystyle -\frac{5z_\mathrm{b}D}{r_\mathrm{b}^7}
\end{bmatrix},
\label{eq:aC22}
\end{equation}
Let $(v_\mathrm{xp},v_\mathrm{yp},v_\mathrm{zp})$ denote body-fixed Cartesian velocities aligned with the principal axes (with $z_\mathrm{b}$ along the spin axis) and $
s_\mathrm{b}\equiv\mathbf{r_\mathrm{b}}\cdot\mathbf{v_\mathrm{b}}=x_\mathrm{b}v_\mathrm{xb}+y_\mathrm{b}v_\mathrm{yb}+z_\mathrm{b}v_\mathrm{zb}$.
The $J_2$ and $C_{22}$ components of the jerk are
\begin{equation}
\mathbf{j}_\mathrm{J2} =
\renewcommand{\arraystretch}{1.5}
\begin{bmatrix}
\dot{k}x_\mathrm{b}A+kv_\mathrm{xb}A+kx_\mathrm{b}\dot{A}\\
\dot{k}y_\mathrm{b}A+kv_\mathrm{yb}A+ky_\mathrm{b}\dot{A}\\
\dot{k}z_\mathrm{b}B+kv_\mathrm{zb }B+kz_\mathrm{b}\dot{B}
\end{bmatrix},
\label{eq:jJ2}
\end{equation}
\begin{equation}
\mathbf{j}_\mathrm{C22} = q
\renewcommand{\arraystretch}{1.5}
\begin{bmatrix}
2v_\mathrm{xp}r_\mathrm{b}^{-5}-10x_\mathrm{b}r^{-7}s_\mathrm{b}-5(v_\mathrm{xb}Dr_\mathrm{b}^{-7}+x_\mathrm{b}\dot{D}r_\mathrm{b}^{-7}-7x_\mathrm{b}Dr_\mathrm{b}^{-9}s_\mathrm{b})\\
-2v_\mathrm{yb}r_\mathrm{b}^{-5}+10y_\mathrm{b}r_\mathrm{b}^{-7}s_\mathrm{b}-5(v_\mathrm{yb}Dr^{-7}+y_\mathrm{b}\dot{D}r_\mathrm{b}^{-7}-7y_\mathrm{b}Dr_\mathrm{b}^{-9}s_\mathrm{b})\\
-5(v_\mathrm{zb}Dr_\mathrm{b}^{-7}+z_\mathrm{b}\dot{D}r_\mathrm{b}^{-7}-7z_\mathrm{b}Dr_\mathrm{b}^{-9}s_\mathrm{b})
\end{bmatrix}.
\label{eq:jC22}
\end{equation}
Additional variables defined in Eqs.~(\ref{eq:aJ2}--\ref{eq:jC22}) are
\begin{equation}
k=\frac{3\mu_\mathrm{b} J_2 R_\mathrm{eq}^2}{2r_\mathrm{b}^5},\quad q=\mu_\mathrm{b} C_{22}R_\mathrm{eq}^2,
\label{eq:aC22}
\end{equation}
\begin{equation}
    \dot{k}=-5k\frac{s_\mathrm{b}}{r_\mathrm{b}^2},
\end{equation}
\begin{equation}
    A=5\frac{z_\mathrm{b}^2}{r_\mathrm{b}^2}-1,\quad B=5\frac{z_\mathrm{b}^2}{r_\mathrm{b}^2}-3,
\end{equation}
\begin{equation}
    \dot{A}=\dot{B}=10\frac{z_\mathrm{b}v_\mathrm{zb}r_\mathrm{b}^2-z_\mathrm{b}^2s_\mathrm{b}}{r_\mathrm{b}^4},
\end{equation}
\begin{equation}
D=x_\mathrm{b}^2-y_\mathrm{b}^2,\quad \dot{D}=2(x_\mathrm{b}v_\mathrm{xb}-y_\mathrm{b}v_\mathrm{yb}).
\end{equation}

\end{document}